\documentclass[intlimits,twoside,a4paper]{article}

\usepackage[cp1251]{inputenc}

\usepackage[eqsecnum]{cmpj3}

\usepackage{bm}
\newcommand{\be}{\begin{equation}}
\newcommand{\ee}{\end{equation}}
\newcommand{\bea}{\begin{eqnarray}}
\newcommand{\eea}{\end{eqnarray}}
\issue{2022}{25}{4}{43711}
\doinumber{10.5488/CMP.25.43711}

\title[Electrocaloric and barocaloric effects]%
{Electrocaloric and barocaloric effects in CsH$_2$PO$_4$ ferroelectric}

\author[A. S. Vdovych, \framebox{R. R. Levitskii}, I. R. Zachek]{ A. S. Vdovych\orcid{0000-0002-1888-8664}\refaddr{label1}\thanks{Corresponding author: vas@icmp.lviv.ua.}, 
\framebox{R. R. Levitskii}\refaddr{label1}, I. R. Zachek\refaddr{label2} }
\addresses{
	\addr{label1} Institute for Condensed Matter Physics of the National Academy of Sciences of Ukraine,\\
	1 Svientsitskii St., 79011 Lviv, Ukraine
	\addr{label2} Lviv Polytechnic National University, 12 Bandera Str., 79013 Lviv, Ukraine 
}


\date{Received June 30, 2022, in final form August 23, 2022}

\begin{document}
	
\maketitle
\begin{abstract}
To investigate  the caloric effects in the CsH$_2$PO$_4$ ferroelectric, a modified pseudospin model of this crystal is used, which
takes into account the dependence of the parameters of interaction between pseudospins on lattice strains. The model also takes into account the dependence of the effective dipole moment of a pseudospin on the order parameter. In the two-particle cluster approximation, the influence of the longitudinal electric field and hydrostatic pressure on the molar entropy of the crystal was studied. The electrocaloric and barocaloric effects were studied. The calculated electrocaloric temperature change is about $1$~K; it can change its sign under the influence of hydrostatic pressure. Barocaloric temperature change is about $-0.5$~K; lattice anharmonicities were not taken into account in its calculations.
\printkeywords
%
%\keywords ferroelectricity, ferroelectric phase transition, electrocaloric effect, barocaloric effect.
%
%\pacs   77.84.-s, 77.22.-d, 77.80.-e, 77.80.Bh, 77.70.+a, 62.50.-p %, 77.65.Bn %, 77.22.Ch
\end{abstract}

\section{Introduction}

Currently, the greatest electrocaloric (EC) effect, as the change in temperature of dielectric with an adiabatic change of electric field, is observed in thin films of perovskite ferroelectrics and relaxors. In particular, there was achieved a change in temperature $\Delta T_{\text{ec}} = 12$~K  in the presence of strong electric field ($E=480$~kV/cm) in crystal PbZr$_{0.95}$Ti$_{0.05}$O$_{3}$ \cite{Mischenko2006},  45.3~K at field strength $E=598$~kV/cm in Pb$_{0.8}$Ba$_{0.2}$ZrO$_{3}$ \cite{Peng2987},   
$-42.5$~K at $E=1632$~kV/cm in 0.5(Ba$_{0.8}$Ca$_{0.2}$)TiO$_{3}$–0.5Bi(Mg$_{0.5}$Ti$_{0.5}$)O$_{3}$ \cite{Peng1708},   
40~K at $E=1200$~kV/cm in Pb$_{0.88}$La$_{0.08}$Zr$_{0.65}$Ti$_{0.35}$O$_{3}$ \cite{Lu162904}. 

In bulk samples, the EC effect is an order of magnitude weaker due to a less dielectric strength.
In particular, there was achieved a temperature change 
$\Delta T_{\text{ec}}=4.5$~K at $E=90$~kV/cm in \linebreak{Pb$_{0.88}$La$_{0.12}$(Zr$_{0.65}$Ti$_{0.35}$)$_{0.97}$O$_3$} \cite{Asbani164517},    
3.5~K at $E=197$~kV/cm in lead scandium tantalate  \cite{Nouchokgwe114873},  
11~K at $E=29.7$~kV/cm in [(CH$_3$)$_2$CHCH$_2$NH$_3$]$_2$PbCl$_4$  \cite{Liu2021}.

In cheaper and more accessible  KH$_{2}$PO$_{4}$ (KDP) type ferroelectrics with hydrogen bonds, the EC effect has been investigated in relatively weak fields or not at all. In particular, in the KDP crystal there was achieved $\Delta T_{\text{ec}} \thickapprox$ 0.04~K at the field strength $E$ $\thickapprox 4$ kV/cm \cite{363x},    $\Delta T_{\text{ec}} \thickapprox$ 1K at $E$ $\thickapprox 12 $~kV/cm \cite{Baumgartner1950}  and $\Delta T \thickapprox$ 0.25~K at temperature $T_c$ at $E$ $\thickapprox 1.2$ kV/cm \cite{Shimshoni1970}.
Calculations carried out in \cite{Vdovych_CMP2014_el} based on the pseudospin model of a deformed KDP crystal show that the $\Delta T_{\text{ec}}$ in this crystal can exceed 5~K.

Ferroelectrics, in which $\Delta T_{\text{ec}}$ is smaller than those mentioned above, are also promising for electrocaloric cooling since, in order to obtaine a given $\Delta T_{\text{ec}}$, electrocaloric devices can be combined into a cascade of several links, in which the heater for the previous link is at the same time the cooler for the next link~\cite{Meng2021}.

In ferroelectric materials, the phase transition temperature depends on the pressure. Therefore, they also exhibit a significant barocaloric (BC) effect, which is a change in the crystal temperature during an adiabatic change in hydrostatic pressure. The strongest BC effect was achieved in crystals with hydrogen bonds NH$_4$HSO$_4$ \cite{Gorev2019} ($\Delta T_{\text{bc}}=-10$~K at pressure $p=0.15$~GPa) and (NH$_4$)$_2$SO$_4$ \cite{Lloveras2015} ($\Delta T_{\text{bc}}=-8$~K at pressure $p=0.1$~GPa).

\begin{figure}[!t]
	\begin{center}
		\includegraphics[scale=0.52]{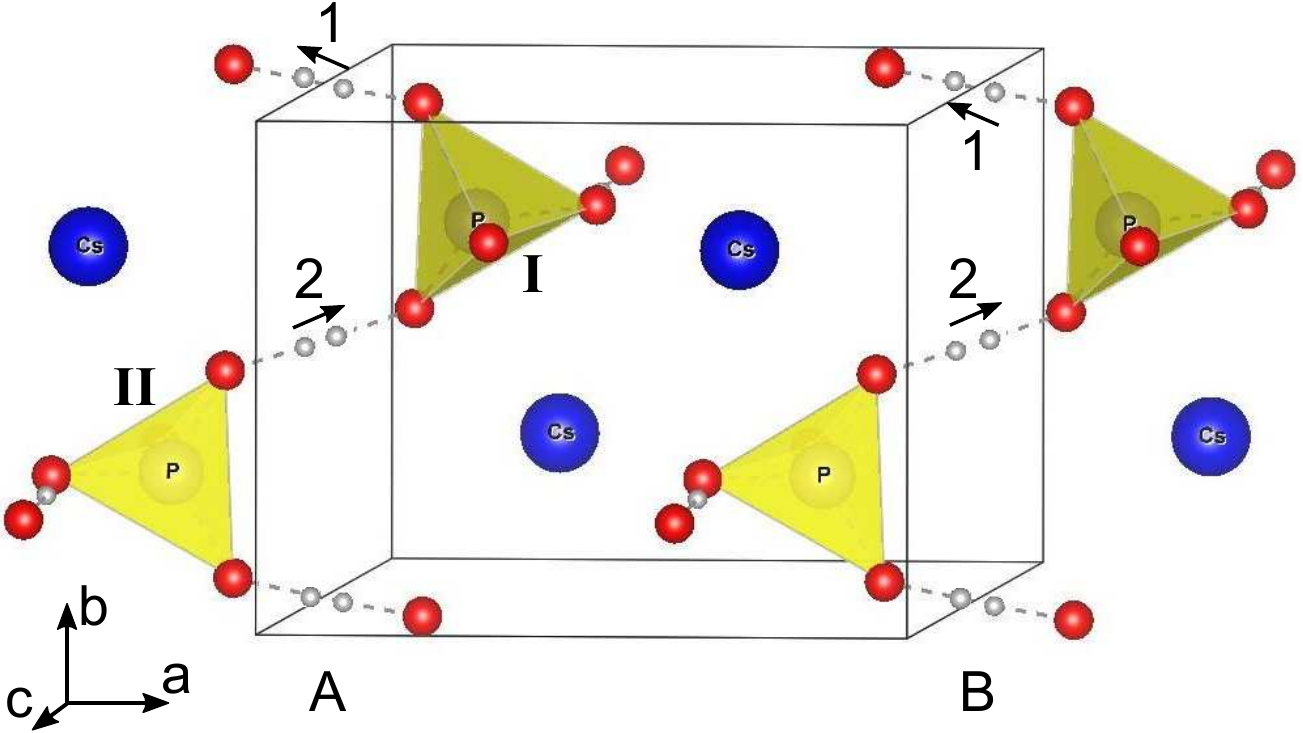}~~~~~~~~\includegraphics[scale=0.52]{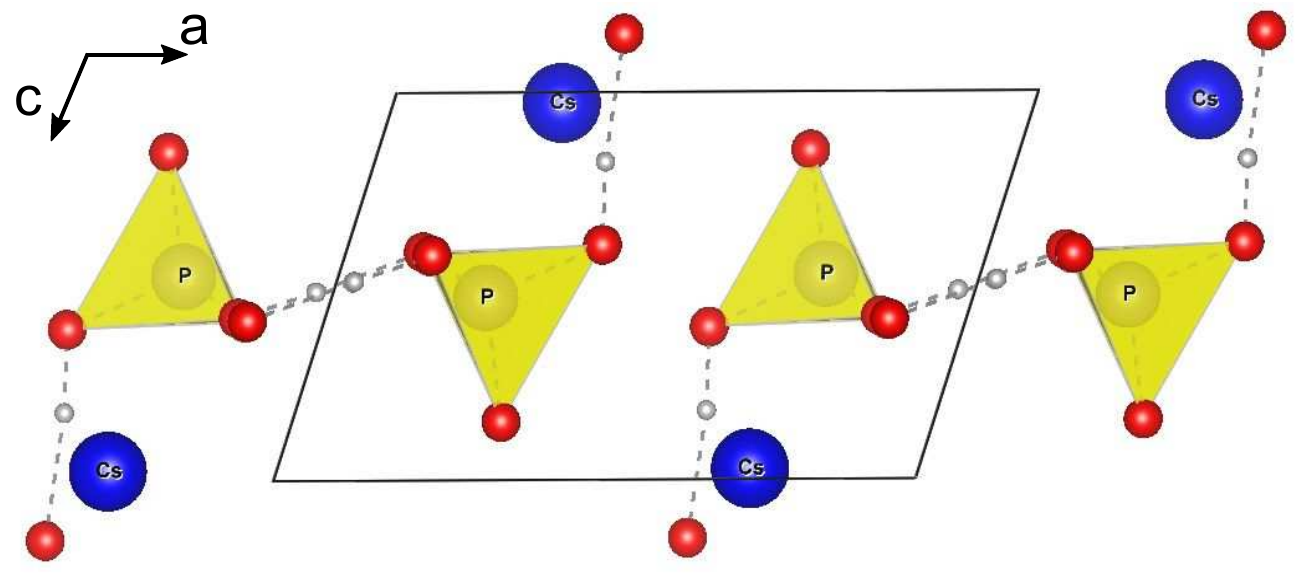} \\
		a ~~~~~~~~~~~~~~~~~~~~~~~~~~~~~~~~~~~~~~~~~~~~~~~~~~~~~~~~~~~~~		b
	\end{center}
	\caption{(Colour online) Primitive cell of CDP crystal in the ferroelectric phase \cite{Vdovych33702}.} 
	\label{CDP_ferro_ab}
\end{figure}

Crystal CsH$_2$PO$_4$ (CDP) is another example of a hydrogen-bonded ferroelectric of the KDP family. Neither EC nor BC effects in this crystal have been studied at all.
In the CDP crystal, there are two structurally non-equivalent types of hydrogen bonds of different lengths (figure~\ref{CDP_ferro_ab}b). Longer bonds have one equilibrium position for protons, while shorter bonds have two equilibrium positions. They connect PO$_4$ groups in chains along the $b$-axis (figure~\ref{CDP_ferro_ab}a); therefore, the crystal is quasi-one-dimensional.

At room temperature in the absence of pressure, the crystal is in the paraelectric phase and has monoclinic symmetry (space group P2$_{1}$/m) \cite{Matsunaga2011,Itoh2626}. At the same time, protons on short bonds are in two equilibrium positions with the same probability. Below $T_{c}=153$~K, the crystal passes to the ferroelectric phase (space group P2$_{1}$) \cite{Iwata304,Iwata4044} with spontaneous polarization along the crystallographic $b$-axis, and protons with a higher probability occupy the upper position (figure~\ref{CDP_ferro_ab}a).
On the basis of dielectric studies~\cite{Yasuda1311,Yasuda2755} 
it was established that at pressures $p=p_{c}=0.33$~GPa and $T_{c}^{\text{cr}}=124.6$~K, double hysteresis loops appear, that is, a transition to the antiferroelectric phase occurs. With the help of neutron diffraction studies \cite{Schuele935}, it was established that in the antiferroelectric phase, the unit cell of the CDP crystal doubles along the \textit{a}-axis, as two sublattices in the form of \textit{bc} planes arise, which are polarized antiparallel along $b $-axis and alternate along the \textit{a}-axis. The symmetry remains monoclinic (space group P2$_{1}$).
Protons on hydrogen bonds are arranged in neighboring sublattices in an antiparallel manner.
At very high pressures, an antiferroelectric phase of the second type (AF2) occurs, in which two sublattices have the form of chains along the \textit{b}-axis, and they are polarized antiparallel along the $b$-axis and alternate in a checkerboard pattern. The AF2 phase was predicted on the basis of NMR studies \cite{Schuele2549} and confirmed in \cite{Kobayashi83} on the basis of X-ray diffraction measurements and dielectric measurements \cite{Deguchi024106}.

The effect of hydrostatic pressure on the phase transition temperature and dielectric properties of Cs(H$_{1-x}$D$_x)_2$PO$_4$ ferroelectrics was studied in  \cite{Gesi,Yasuda1311,Yasuda2755,Brandt,Kobayashi83,Magome2010,Deguchi024106}.
The molar heat capacity of CDP was measured in \cite{Imai3960}, and was also calculated based on the lattice dynamics simulations in \cite{Shchur054301,Shchur569}.
Later, based on the ab-initio calculations \cite{Lasave134112} and using calculations based on the quasi-one-dimensional model \cite{Kojyo4391}, the important role of proton tunneling on the bonds was established. Piezoelectric coefficients, elastic constants, and molar heat capacity of CDP \cite{Shchur301,VanTroeye024112} were also calculated on the basis of first-principle calculations.

A theoretical description of the dielectric properties of CDP at different values of hydrostatic pressure was carried out in \cite{Blinc6031,914R} based on the pseudospin model.
However, in these works, the interaction parameters do not depend on the lattice strains. As a result, it is impossible to obtain piezoelectric and elastic characteristics of the crystal, and the critical pressure does not depend on temperature.

In \cite{Deguchi3074}, temperature dependences of lattice strains $u_1$, $u_2$, $u_3$, $u_5$ were measured. A quasi-one-dimensional Ising model for the CDP crystal is also proposed there, in which the interaction parameters are linear functions of these strains. Based on this model, the temperature behavior of $u_j(T)$ was explained. However, this model does not consider the crystal as two sublattices and does not allow describing the ferro-antiferroelectric transition at high pressures.

In the papers \cite{FXTT40,Levitskii4702,Vdovych33702,Vdovych_JPS2021_CDP}, a two-sublattice pseudospin model of a deformed CDP crystal is proposed, in which the interactions between the nearest pseudospins in the chain are taken into account in the two-particle cluster approximation, and the long-range (including interchain) interactions are taken into account in the mean field approximation. At the same time, the interaction parameters are linear functions of $u_j$ strains. As a result, the temperature dependences of spontaneous polarization, dielectric constant, piezoelectric coefficients and elastic constants were calculated, and the influence of hydrostatic and uniaxial pressures and longitudinal electric field on these characteristics was studied.
In \cite{Vdovych_JPS2021_CDP}, the behavior of the thermodynamic characteristics of the CDP crystal under the action of hydrostatic and uniaxial pressures and a longitudinal electric field, as well as under the simultaneous action of pressures and the electric field, was investigated.

In the present paper, the electrocaloric and barocaloric effects in CDP crystal are calculated based on the model proposed in \cite{Vdovych33702}.

\section{Model of CDP crystal}

The \cite{Vdovych33702} model was used to calculate the thermodynamic characteristics of CDP,
which considers the system of protons on O-H...O bonds with a two-minimum potential as a system of pseudospins.
The primitive cell contains one chain, marked in figure~\ref{CDP_ferro_ab} as ``A''. To describe the transition to the antiferroelectric phase at high pressures, in 
\cite{Vdovych33702} an extended primitive cell formed by two chains (``A'' and ``B'') is considered. All ``A'' chains form the ``A'' sublattice, and all ``B'' chains form the ``B'' sublattice. Each chain in the primitive cell contains two neighboring PO$_4$ tetrahedra (of type ``I'' and ``II'') together with two short hydrogen bonds (``1'' and ``2'', respectively).
The dipole moments ${\vec{d}}_{q1}^A$, ${\vec{d}}_{q2}^A$, ${\vec{d}}_{ q1}^B$, ${\vec{d}}_{q2}^B$ are attributed to the protons on the bonds.
Pseudospin variables
${\sigma_{q1}^A}/{2}$, ${\sigma_{q2}^A}/{2}$, ${\sigma_{q1}^B}/{2}$, ${\sigma_{q2}^B}/{2}$ describe
changes associated with the rearrangement of the corresponding dipole moments of
structural units: ${\vec{d}}_{q1,2}^{A,B} = \vec\mu_{q1,2}^{A,B} \frac{\sigma_{q1,2}^ {A,B}}{2}$.

Further, we  use the notation ``2'' instead of ``y'' for the components of vectors and tensors, for convenience.
In the presence of mechanical stresses that do not change the symmetry of the crystal $\sigma_1 = \sigma_{xx}$, $\sigma_2 = \sigma_{yy}$, $\sigma_3 = \sigma_{zz}$, $\sigma_5 = \sigma_{ xz}$ (X $\perp$ (\textit{b},\textit{c}), Y $\parallel$ \textit{b}, Z $\parallel$ \textit{c}), as well as  of the electric field $E_2=E_y$, the Hamiltonian of the CDP model has the form \cite{Vdovych33702}:
\setcounter{equation}{0}
\renewcommand{\theequation}{2.\arabic{equation}}
\bea
&& \hat H= NU_{\text{seed}} + \hat H_{\text{short}} + \hat H_{\text{long}} + \hat H_{E} + \hat H'_{E}, \label{H_CDP}
\eea
where $N$ is the total number of extended primitive cells.

The first term in (\ref{H_CDP}) is the ``seed'' energy, which corresponds to the lattice of heavy ions and  does not explicitly depend on the configuration of the proton subsystem. It includes elastic, piezoelectric and dielectric parts expressed through
electric field $E_2$ and strains that do not change the lattice symmetry, $u_1=u_{xx}$, $u_2=u_{yy}$, $u_3=u_{zz}$, $u_5=2u_{xz}$:
\bea
&&  \hspace{-4ex} U_{\text{seed}} = v\left\lbrace \frac12 \sum\limits_{j,j'} c_{jj'}^{E0} u_ju_j'   - \sum\limits_{j} e_{2j}^0E_2u_j   - \frac12 \varepsilon_0\chi_{22}^{u 0}E_2^2 \right\rbrace, ~~~~ j, j'=1,2,3,5,  \label{Useed}
\eea
where  $\varepsilon_0 = 8.8542\cdot10^{-12}$~F/m is electric constant,  $c_{jj'}^{E0}$, $e_{2j}^0$, $\chi_{22}^{u 0}$ are ``seed'' elastic constants, piezoelectric stress coefficients and dielectric susceptibility of a mechanically clamped crystal. $v$ is the volume of the extended primitive cell.
In the paraelectric phase, all coefficients $e_{2j}^0 \equiv 0$.

The other terms in (\ref{H_CDP}) describe the pseudospin part of the Hamiltonian.
In particular, the second term in (\ref{H_CDP}) is the Hamiltonian of short-range interactions
\be
\hat H_{\text{short}} = - 2w \sum\limits_{qq'} \left ( \frac{\sigma_{q1}^{A}}{2} \frac{\sigma_{q'2}^{A}}{2} + \frac{\sigma_{q1}^{B}}{2}\frac{\sigma_{q'2}^{B}}{2} \right) \bigl( \delta_{{\bf R}_q{\bf R}_{q'}} + \delta_{{\bf R}_q + {\bf R}_b,{\bf R}_{q'}} \bigr). \label{Hshort}
\ee
In (\ref{Hshort}),  $\sigma_{q1,2}^{A,B}$ are  $z$-components of pseudospin operator, that describe the state of the bond ``1''  or ``2'' of the chain ``A'' or ``B'', in the  $q$-th cell, ${\vec R}_b$ is the lattice vector along $OY$-axis.
The first Kronecker delta corresponds to the interaction between neighboring pseudospins in the chains near the tetrahedra PO$_{4}$ of type ``I'', where the second Kronecker delta is near the tetrahedra PO$_{4}$ of type ``II''. 
Contributions to the energy of interactions between pseudospins near tetrahedra of different types are identical.
Parameter $w$,  which describes the short-range interactions within the chains, is expanded
linearly into a series with respect to strains $u_j$:
\be
w = w_0 + \sum\limits_{j} \delta_{j}u_j,(j=1,2,3,5). \label{w}
\ee

The  term $\hat H_{\text{long}}$ in  (\ref{H_CDP}) describes  long-range dipole-dipole interactions and indirect  (through the lattice vibrations)  interactions between pseudospins which are taken into account in the mean field approximation:
\be
\hat H_{\text{long}} = N H^0 + \hat H_2, \label{Hlongg}
\ee
where such notations are used:
\bea
&&  \hspace{-8ex}\hat H^{0} = \nu_{1}( \eta_{1}^2 + \eta_{2}^2) + 2\nu_{2}\eta_{1}\eta_{2}, \label{H0}\\
%\eea
%%
%\bea
&&  \hspace{-8ex} \hat H_{2} = \sum\limits_{q} \left\{- (2\nu_{1}\eta_{1} +  2 \nu_{2}\eta_{2} )  \! \left( \!\frac{\sigma_{q1}^{A}}{2} + \frac{\sigma_{q2}^{A}}{2} \!\right)\! - (2\nu_{2}\eta_{1} +  2 \nu_{1}\eta_{2} )  \! \left(\! \frac{\sigma_{q1}^{B}}{2} + \frac{\sigma_{q2}^{B}}{2} \!\right)\right\}.  \label{H2} \\
%\eea
%
%\bea
&&  \hspace{-8ex} \nu_{1} = \nu_{1}^0 + \sum\limits_{j}  \psi_{j1}u_j,~  \nu_{2} = \nu_{2}^0 + \sum\limits_{j}  \psi_{j2}u_j, ~~~~ \langle \sigma_{q1}^{A} \rangle = \langle \sigma_{q2}^{A} \rangle = \eta_{1}, \quad
\langle \sigma_{q1}^{B} \rangle = \langle \sigma_{q2}^{B} \rangle = \eta_{2}. \label{nu}
\eea
The parameter $\nu_{1}$ describes the effective long-range interaction of the pseudospin with the pseudospins within the same sublattice, and $\nu_{2}$ --- with the pseudospins of the other sublattice.

The fourth term in  (\ref{H_CDP}) describes the interactions of pseudospins with the external electric field:
\bea
&&\hat H_{E}= - \sum\limits_q \mu_y E_2 \left( \frac{\sigma_{q1}^{A}}{2} + \frac{\sigma_{q2}^{A}}{2} + \frac{\sigma_{q1}^{B}}{2} + \frac{\sigma_{q2}^{B}}{2} \right),
\eea
where $\mu_y$ is \textit{y}-component of effective dipole moments per one pseudospin.

The term $\hat H'_{E}$ in Hamiltonian (\ref{H_CDP}) takes into account the dependence of the effective dipole moment on the mean value of pseudospin $s_f$:
\bea
&&\hat H'_{E} = -\sum\limits_{qf} s_f^2 \mu' E_2 \frac{\sigma_{qf}}{2} = - \sum\limits_{qf} \left(\frac{1}{N}\sum\limits_{q'}\sigma_{q'f}\right)^2 \mu' E_2 \frac{\sigma_{qf}}{2}, \label{H_E}
\eea
where $\sigma_{qf}$ (\textit{f}=1, 2, 3, 4) are a brief notation of pseudospins $\sigma_{q1}^{A}$, $\sigma_{q2}^{A}$, $\sigma_{q1}^{B}$, $\sigma_{q2}^{B}$, respectively. 
Here, we use corrections to dipole moments $s_f^2 \mu'$ instead of $s_f \mu'$ because of the symmetry considerations and the energy should not change when the field and all pseudospins change their sign.

The term $\hat H'_E$, as well as long-range interactions, is taken into account in the mean field approximation: 
\bea
%&& \hspace{-2ex}  \hat H'_{E} =   -12 \sum\limits_{qf}(\eta_1+\eta_2)^2 \mu'E_2 \frac{\sigma_{qf}}{2}  + 16N(\eta_1+\eta_2)^3 \mu'E_2.
&& \hspace{-8ex}  \hat H'_{E} =   -3 \sum\limits_{q} \mu'E_2 \left( \frac{\eta_1^2\sigma_{q1}^{A}}{2} + \frac{\eta_1^2\sigma_{q2}^{A}}{2} + \frac{\eta_2^2\sigma_{q1}^{B}}{2} + \frac{\eta_2^2\sigma_{q2}^{B}}{2} \right)  + 2N(\eta_1^3+\eta_2^3) \mu'E_2.
\eea
In the two-particle cluster approximation for short-range interactions, the thermodynamic potential per one extended primitive cell is as follows:
\bea
 g &=& U_{\text{seed}} + H^0  + 2(\eta_1^3+\eta_2^3) \mu'E_2 + 2k_\text{B}T \ln 2 - 2w - v \sum\limits_{j} \sigma_j u_j \nonumber \\
&-&k_\text{B}T\ln ( 1 -  \eta_{1}^2) - k_\text{B}T\ln ( 1 -  \eta_{2}^2 )  - 2k_\text{B}T\ln D.
\label{g}
\eea
Here, the following notations are used:
\bea
&& D = \cosh (y_{1} + y_{2}) +  \cosh (y_{1} - y_{2}) + 2a \cosh y_{1}+ 2a \cosh y_{2}  + 2a^2,  ~~~~ a = \re^{-\beta w}.\nonumber \\
%\eea
%%
%\bea
&& y_{1} = \frac12 \ln \frac{1 +  \eta_{1}}{1 - \eta_{1}}  + \beta \nu_{1}\eta_{1}  +   \beta \nu_{2} \eta_{2} + \frac{1}{2}\beta(\mu_yE_2 + 3\eta_1^2 \mu'E_2), \nonumber \label{y1x}\\
&& y_{2} = \frac12 \ln \frac{1 +  \eta_{2}}{1 - \eta_{2}}  + \beta \nu_{2}\eta_{1}  +   \beta \nu_{1} \eta_{2}+ \frac{1}{2}\beta(\mu_yE_2 + 3\eta_2^2 \mu'E_2), \nonumber \label{y2x}
\eea
where $\beta=\frac{1}{k_\text{B}T}$, $k_\text{B}$ is Boltzmann constant.

Minimizing the thermodynamic potential with respect to the order parameters $\eta_{f}$ and strains $u_j$ in~\cite{Vdovych33702}, we obtain a system of equations for $\eta_{f}$ and $u_j$:
\bea
&& \eta_{1} = \frac{1}{D} \left[ \sinh (y_{1} + y_{2}) +  \sinh (y_{1} - y_{2}) + 2a \sinh y_{1}\right], \label{eta1eta2} \\
&& \eta_{2} = \frac{1}{D} \left[ \sinh (y_{1} + y_{2}) -  \sinh (y_{1} - y_{2}) + 2a \sinh y_{2} \right], \nonumber \\
%\eea
%
%\bea
&& \sigma_j = c_{j1}^{E0}u_1 + c_{j2}^{E0}u_2 + c_{j3}^{E0}u_3 + c_{j5}^{E0}u_5 - e_{2j}^0E_2 - \frac{2\delta_j}{v} + \frac{4\delta_j}{vD}M  - \frac{1}{v} \psi_{j1} ( \eta_{1}^{2}  +  \eta_{2}^{2}  ) - \frac{2}{v} \psi_{j2} \eta_{1}\eta_{2}, \nonumber \label{sigma} 
%&& \sigma_4 = c_{44}^{E0}u_4 + c_{46}^{E0}u_6 - e_{14}^0E_1 - e_{16}^0E_3   - \frac{1}{v} \psi_{41s,a} \eta_{1s,a}^{2}+ \frac{1}{v} \psi_{41s,a} \eta_{2s,a}^{2}, \nonumber\\
%&& \sigma_6 = c_{46}^{E0}u_4 + c_{66}^{E0}u_6 - e_{34}^0E_1 - e_{36}^0E_3   - \frac{1}{v} \psi_{61s,a} \eta_{1s,a}^{2} + \frac{1}{v} \psi_{61s,a} \eta_{2s,a}^{2},  \nonumber\\
% && P_1 =  e_{14}^0u_4 + e_{16}^0u_4 + \chi_{11}^{u 0}E_1 + \chi_{13}^{u 0}E_3 + \frac{\mu_x}{v} \bigl(  \eta_{1s,a} - \eta_{2s,a} \bigr), \\
\eea
where
\[ M=\bigl[   a\cosh y_{1} + a\cosh y_{2} + 2a^{2} \bigr]. \]
In the presence of hydrostatic pressure $\sigma_1=\sigma_2=\sigma_3=-p$,  $\sigma_4=\sigma_5=\sigma_6=0$.

In \cite{Vdovych33702}, the expression for the longitudinal component of polarization $P_2$ was also obtained:
\bea
&& \hspace{-4ex} P_2 = -\left( \frac{\partial g}{\partial E_2} \right)_{\sigma_j}   =   \sum\limits_{j} e_{2j}^0u_j   + \chi_{22}^{u 0}E_2  + \frac{\mu_y}{v} \bigl(  \eta_{1} + \eta_{2} \bigr)  +  \frac{\mu'}{v} \bigl(  \eta_{1}^3 + \eta_{2}^3 \bigr). \label{P2}
\eea
%\bea
%&& \hspace{-4ex} P_2 =  \sum\limits_{j} e_{2j}^0u_j   + \chi_{22}^{u 0}E_2  + \frac{2\mu_y}{v} \eta  +  \frac{2\mu'}{v} \eta^3. \label{P2}
%\eea

Based on the thermodynamic potential (\ref{g}), we obtain an expression for the entropy of the pseudospin subsystem:%
\bea
S = &-& \frac{N_A}{N_m} \left(   \frac{\partial g}{\partial T} \right)_{\eta,\varepsilon_i} 
= \frac{R}{N_m} \left\lbrace  - 2\ln2 + \sum\limits_{f=1}^2\ln\bigl( 1 - \eta_{f}^2 \bigr)+ 2\ln D  \right. \nonumber \\
& -& 2\eta_1 \beta \left[ \nu_{1}\eta_{1}  +   \nu_{2} \eta_{2} + \frac{1}{2}\left( \mu_yE_2 + 3\eta_1^2 \mu'E_2\right) \right]   \nonumber \\
 &-& \left. 2\eta_2 \beta \left[ \nu_{2}\eta_{1}  +   \nu_{1} \eta_{2} + \frac{1}{2}\left( \mu_yE_2 + 3\eta_2^2 \mu'E_2\right) \right]  +  \frac{4M\beta w}{D}\right\rbrace .
 \label{S}
\eea
Here, $N_\text{A}$ is Avogadro constant, $R$ is the universal gas constant, $N_m=4$ is the number of CsH$_{2}$PO$_{4}$ molecules in the extended primitive cell.

The molar heat capacity of the pseudospin subsystem of the CDP crystal:%
\bea
&& \hspace{-8ex} C = T  \left( \frac{\rd S}{\rd T} \right)_{E_2,\sigma_j} = T\left(  S'_{T} + \sum\limits_{f=1}^2 S'_{\eta_f} \eta'_{Tf}  + \!\! \sum\limits_{j=1,2,3,5} \!\! S'_{u_j} u'_{Tj} \right) .  \label{DC}
\eea
The explicit expressions for derivatives $S'_{T}$, $S'_{\eta_f}$, $S'_{u_j}$, $\eta'_{Tf}$, $u'_{Tj}$ are given in the appendix.

We consider the total heat capacity to be the sum of the pseudospin and lattice components:
\be C_{\text{total}}=C + C_{\text{lattice}} . \ee
The heat capacity of the lattice subsystem is considered to be the CDP heat capacity, calculated on the basis of first-principle calculations \cite{Shchur301}. Its temperature dependence in the range of 80--350~K, in which the calculations were carried out, is well approximated by a polynomial
\be C_{\text{lattice}} = \sum\limits_{l=0}^4 k_l T^l, \label{Clattice}\ee
where the coefficients $k_l$: $k_0=17.62$~J/(mol K), $k_1=0.5955$~J/(mol K$^2$), $k_2=-0.001885$~J/(mol K$^3$), $k_3=4.376\cdot$ 10$^{-6}$~J/(mol K$^4$), $k_4=-4.034\cdot$ $10^{-9}$~J/(mol K$^5$).
The entropy of the lattice subsystem near $ T_c $:
\be S_{\text{lattice}} = \int \frac{C_{\text{lattice}}}{T}\rd T = k_0 \ln(T)+\sum\limits_{l=1}^4 \frac{k_l T^l}{l} + \text{const}. \label{Slattice}\ee
Total entropy as a function of temperature, field component $E_2$ and hydrostatic pressure $p$:
\be S_{\text{total}}(T,E_2,p)=S + S_{\text{lattice}}. 
\label{Stotal} \ee
Solving (\ref{Stotal}) with respect to the temperature at $S_{\text{total}}(T,E_2,p)=\text{const}$ and two magnitudes of the field, it is possible to calculate the electrocaloric temperature change (as shown in figure~\ref{S_E_p}b):
\be \Delta T_{\text{ec}} = T[S_{\text{total}},E_2(2),p]-T[S_{\text{total}},E_2(1),p]. \label{DT_ek} \ee

The change in temperature during the adiabatic change in the field $E_2$ can also be calculated by the well-known formula
\be \Delta T_{\text{ec}} = -\int \limits_{0}^{E_2} \frac{TV}{C_{\text{total}}} \left(\frac{\partial P_2}{\partial T}
\right)_{E_2}\rd E_2,  \label{DTec_int}\ee
where pyroelectric coefficient
\bea \left(\frac{\partial P_2}{\partial T} \right)_{E_2} = 
\sum\limits_{j} e_{2j}^0u'_{jT}   + \frac{\mu_y}{v} \bigl(  \eta'_{1T} + \eta'_{2T} \bigr)  +  \frac{3\mu'}{v} \bigl(  \eta_{1}^2\eta'_{1T} + \eta_{2}^2\eta'_{2T} \bigr), \label{P2T}
\eea
and $V=vN_A/N_m$ is molar volume.

Similarly, solving (\ref{Stotal}) with respect to temperature at $S_{\text{total}}(T,E_2,p)=\text{const}$ and two pressure values, it is possible to calculate the barocaloric temperature change (as shown in figure~\ref{S_E_p}b):
\be \Delta T_{\text{bc}} = T[S_{\text{total}},E_2,p(2)]-T[S_{\text{total}},E_2,p(1)]. \label{DT_bk} \ee

The change in temperature under the adiabatic change in pressure $p$ can also be calculated by the known formula
\be \Delta T_{\text{bc}} = \int \limits_{0}^{p} \frac{T}{C_{\text{total}}} \left(\frac{\partial V}{\partial T}
\right)_{p}\rd p  =  \int \limits_{0}^{p} \frac{N_AT}{N_mC_{\text{total}}} (u'_{1T}+u'_{2T}+u'_{3T})\rd p.  \label{DTbc_int}\ee

\section{Discussion of the obtained results}

The theory parameters are  determined in \cite{Vdovych33702} from the condition of agreement of calculated characteristics with experimental data for temperature dependences of spontaneous polarization $P_2(T)$ and dielectric permittivity  $\varepsilon_{22}(T)$ at different values of hydrostatic pressure  \cite{Yasuda2755}, spontaneous strains $u_j$ \cite{Deguchi3074}, molar heat capacity \cite{Imai3960} and elastic constants  \cite{Prawer63};  as well as agreement with  ab-initio calculations of the  lattice contributions into molar heat capacity  \cite{Shchur301} and dielectric permittivity at zero temperature \cite{VanTroeye024112}.  

It should be noted that the temperature dependences of the dielectric constant $\varepsilon_{22}$ at different values of hydrostatic pressure were also measured in \cite{Deguchi024106}. However, they do not agree with experimental data \cite{Yasuda2755}. It is possible that another crystal sample was used there, which was grown under different conditions. In addition, in \cite{Deguchi024106} there are no data for the temperature dependences of spontaneous polarization at different pressures, as well as no data for dielectric characteristics at zero pressure. Therefore, we used experimental data \cite{Yasuda2755} to determine the model parameters.

Parameters of short-range interactions $w_0$ and long-range interactions  $\nu_1^{0}$ (``intra-sublattice''), $\nu_2^{0}$ (``inter-sublattice'') mainly fix the phase transition temperature from  paraelectric to ferroelectric phase at the absence of external pressure and field, the order of  phase transition and the shape of curve $P_2(T)$. Their optimal values are: $w_0/k_{\text{B}}=650$~K,  $\nu_1^{0}/k_{\text{B}}=1.50$~K,  $\nu_2^{0}/k_{\text{B}}=0.23$~K.

In order to determine the deformational potentials $\delta_{j}$ [see (\ref{w})] and $\psi_{j1}$, $\psi_{j2}$ [see (\ref{nu})], it is necessary to use experimental data for the shift of the phase transition temperature under hydrostatic and uniaxial pressures as well as the data for temperature dependences of  spontaneous strains $u_j$, piezoelectric coefficients and elastic constants. Unfortunately,  only the data for the spontaneous strains and hydrostatic pressure effect on the dielectric characteristics are available. 
As a result,  the experimental data for strains and dielectric characteristics can be described using a great number of combinations of parameters $\psi_{j1}$,  $\psi_{j2}$. 
Therefore, for the sake of simplicity, we chose $\psi_{j2}$ to be proportional to $\psi_{j1}$.  
Optimal values of deformational potentials are: 
$\delta_{1}/k_{\text{B}}=1214$~K, $\delta_{2}/k_{\text{B}}=454$~K, $\delta_{3}/k_{\text{B}}=1728$~K, $\delta_{5}/k_{\text{B}}=-131$~K;
$\psi_{11}/k_{\text{B}} = 92.2$~K,  $\psi_{21}/k_{\text{B}} = 23.2$~K,  $\psi_{31}/k_{\text{B}} = 139.7$~K,  $\psi_{51}/k_{\text{B}} = 5.5$~K;
$\psi_{j2}$ = $\frac{1}{3}\psi_{j1}$.

The effective dipole moment in the paraelectric phase is found from the condition of agreement of the calculated curve $\varepsilon_{22}(T)$  with experimental data. We consider it to be dependent on the value of hydrostatic pressure $p$, that is $\mu_{y}=\mu_{y}^0(1-k_pp)$, where $\mu_{y}^0=8.77\cdot 10^{-30}$~C$\cdot$m,  $k_p=0.4\cdot 10^{-9}$~Pa$^{-1}$.
The correction to the effective dipole moment $\mu'=-1.43\cdot 10^{-30}$~C$\cdot$m  is found from the condition of agreement of the calculated saturation polarization with experimental data.

The ``seed'' dielectric susceptibility  $\chi_{22}^{u 0}$, coefficients of piezoelectric stress $e_{2j}^0$ and elastic constants $c_{ij}^{E0}$  are found from the condition of agreement of theory with experimental data in the temperature regions far from the phase transition temperature  $T_c$. Their values are obtained as follows:
$\chi_{22}^{u 0}=5.57$;   $e_{2j}^0=0$~C/m$^2$;   $c_{jj'}^{E0}$ ($10^{9}$N/m$^2$):
$c_{11}^{E0}=28.83$, $c_{12}^{E0}=11.4$, $c_{13}^{E0}=42.87$, $c_{22}^{E0}=26.67$, $c_{23}^{E0}=14.5$, $c_{33}^{E0}=65.45$, $c_{15}^{E0}=5.13$,  $c_{25}^{E0}=8.4$, $c_{35}^{E0}=7.50$, 
$c_{55}^{E0}=5.20$.

The volume of the extended primitive cell is  $\upsilon = 0.467\cdot 10^{-27}$ m$^3$ \cite{Schuele935}.

\begin{figure}[!h]
	\begin{center}
		\includegraphics[scale=0.85]{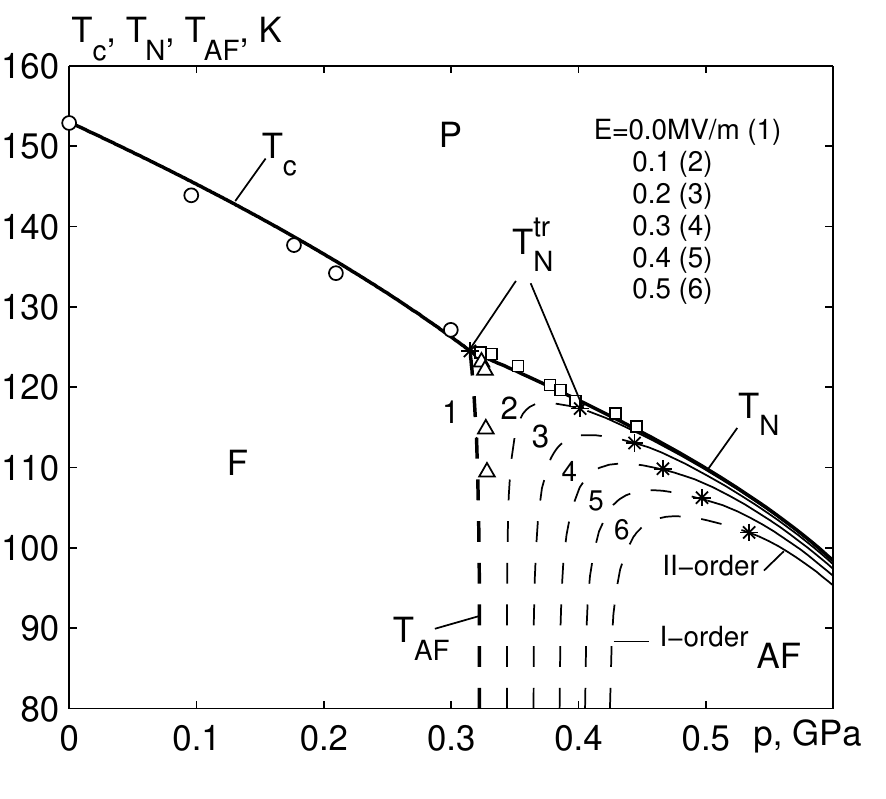} 
	\end{center}
	\caption{Dependence on the hydrostatic pressure of the temperature of the transition from the paraelectric to the ferroelectric phase $T_c$, from the paraelectric  to the antiferroelectric phase $T_N$, from the ferroelectric  to the antiferroelectric phase $T_{AF}$ at different values of the electric field $E_2$ (MV/m): 0.0 --1 , 0.1 -- 2, 0.2 -- 3, 0.3 -- 4, 0.4 -- 5, 0.5 -- 6 for the CDP crystal. Symbols are experimental data \cite{Yasuda1311}, lines are theoretical calculations \cite{Vdovych33702}. Tricritical points $ T_N^{\text{tr}} $ (marked as *) separate the curves of the first-order phase transitions (dashed lines) and of the second-order ones (solid lines).
	} \label{TcTN_E2}
\end{figure}

In the paper \cite{Vdovych33702}, a phase diagram (figure~\ref{TcTN_E2}) was calculated, which explains the effect of hydrostatic pressure and longitudinal electric field on the temperatures of phase transitions, in particular, the transition to the antiferroelectric phase at pressures greater than the critical one.

\begin{figure}[!t]
	\begin{center}
		\includegraphics[scale=0.9]{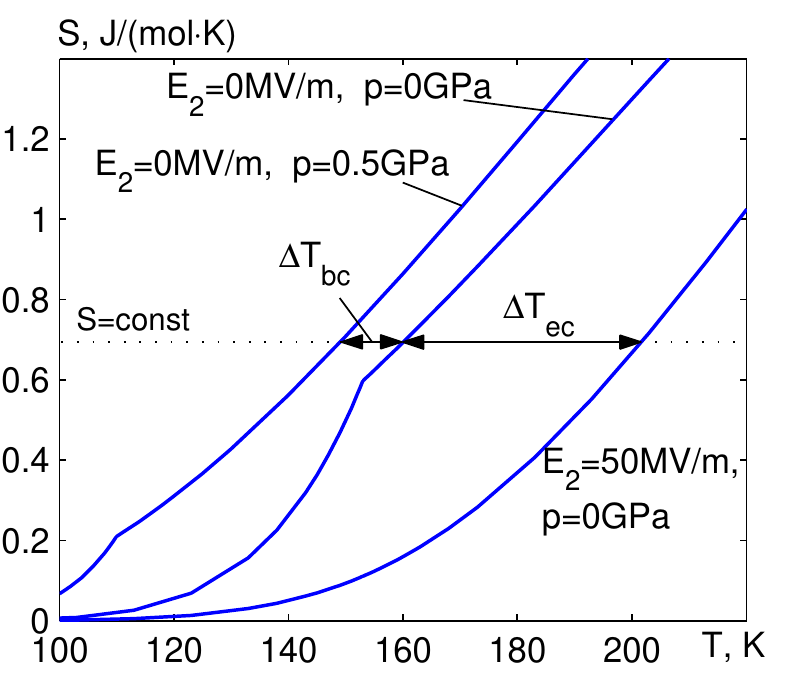} \includegraphics[scale=0.9]{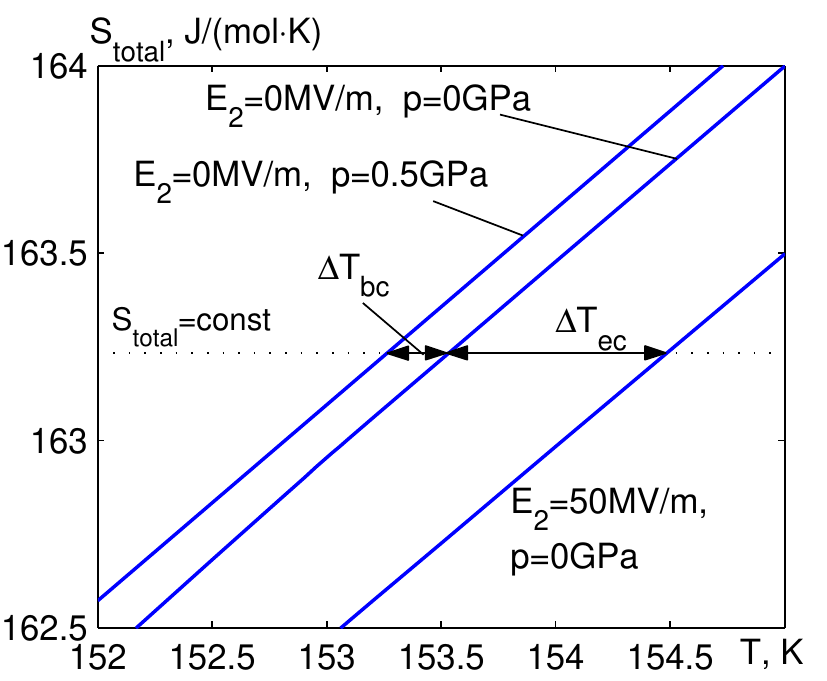} \\
		a ~~~~~~~~~~~~~~~~~~~~~~~~~~~~~~~~~~~~~~~~~~~~~~~~~~~~~~~~~~~~~~~ b
	\end{center}
	\caption{(Colour online) Temperature dependences of the pseudospin contribution to the molar entropy~(a) and total entropy (b) of the CDP crystal at different values of the field $E_2$ and of the hydrostatic pressure~$p$.} 
	\label{S_E_p}
\end{figure}

As mentioned above, the EC effect is calculated as a change in the crystal temperature $\Delta T_{\text{ec}}$ during adiabatic (at constant entropy) application of an electric field, as shown in figure~\ref{S_E_p}.
At the pressures less than critical,  longitudinal field $E_2$ decreases the entropy of the crystal in the entire temperature range (figure~\ref{S_E_p}), because it puts the pseudospins in order  in both sublattices, ``A'' and ``B'' (figure~\ref{CDP_ferro_ab}a). Therefore, the $\Delta T_{\text{ec}}$ is positive.
As we can see, the effect of the field on the total entropy $S_{\text{total}}$ (figure~\ref{S_E_p}b) is much weaker than the effect on only the pseudospin contribution $S$ (figure~\ref{S_E_p}a), because the lattice heat capacity quite strongly stabilizes the temperature of the crystal.

\begin{figure}[!h]
	\begin{center}
		\includegraphics[scale=0.9]{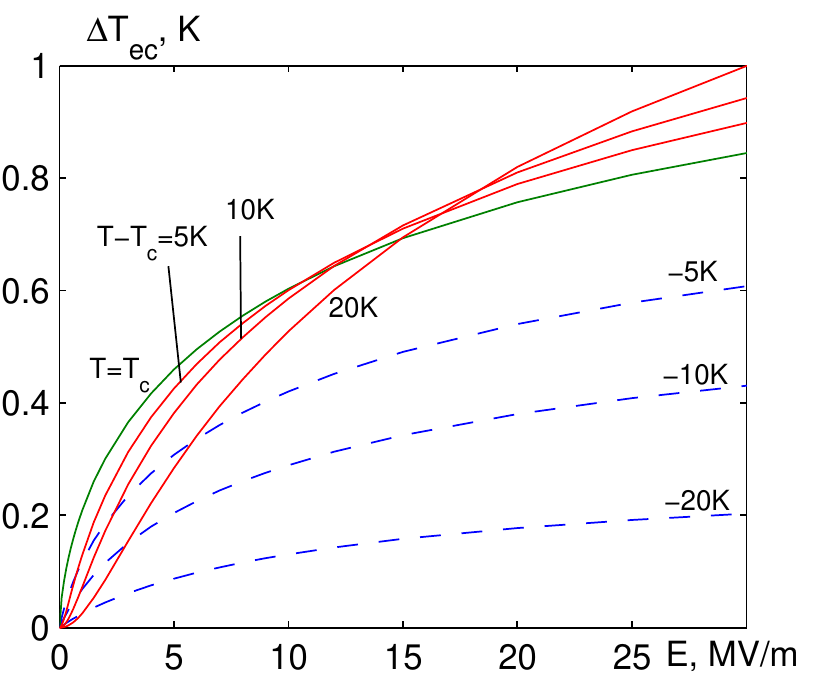} \includegraphics[scale=0.9]{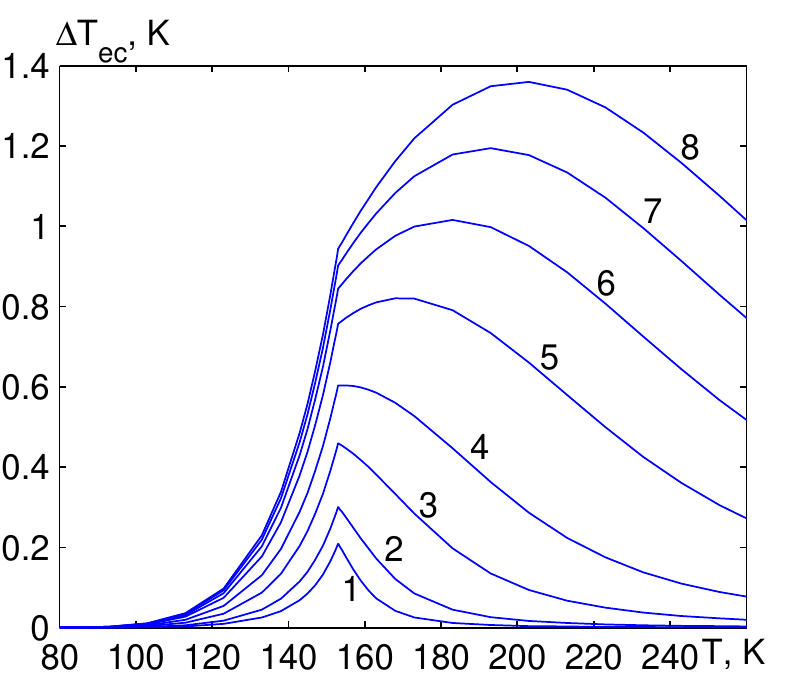} \\
		a ~~~~~~~~~~~~~~~~~~~~~~~~~~~~~~~~~~~~~~~~~~~~~~~~~~~~~~~~~~~~~~~ b
	\end{center}
	\caption{(Colour online) a) Field dependence of the electrocaloric temperature change $\Delta T_{\text{ec}}$ at different values of temperature $\Delta T = T-T_c$ and at zero hydrostatic pressure $p$. ~b) Temperature dependence of $\Delta T_{\text{ec}}$ at different values of the longitudinal electric field $E_2$ (MV/m): 1.0 -- 1; 2.0 -- 2; 5.0 -- 3; 10.0 -- 4; 20.0 -- 5; 30.0 -- 6; 40.0 -- 7; 50.0 -- 8 and at zero hydrostatic pressure $p$. } 
	\label{DTec_E_0kbar}
\end{figure}

The calculated field and temperature dependences of  $\Delta T_{\text{ec}}$ are shown in figure~\ref{DTec_E_0kbar}.
In the weak fields ($E_2<1$~MV/m) at the initial temperature $T=T_c$, the change in temperature $\Delta T_{\text{ec}}\sim E_2^{2/3}$ (green curve in figure~\ref{DTec_E_0kbar}a); at $T<T_c$, $\Delta T_{\text{ec}}\sim E_2$ (blue dashed curves in figure~\ref{DTec_E_0kbar}); at \textit{T}>$T_c$, $\Delta T_{\text{ec}}\sim E_2^2$ (red curves in figure~\ref{DTec_E_0kbar}). At fields $E_2>1$~MV/m, the dependences of $\Delta T_{\text{ec}}(E_2)$ significantly deviate from the mentioned laws.
%, and at $E_2\gg50$MV/m they reach saturation.

\begin{figure}[!t]
	\begin{center}
		\includegraphics[scale=0.85]{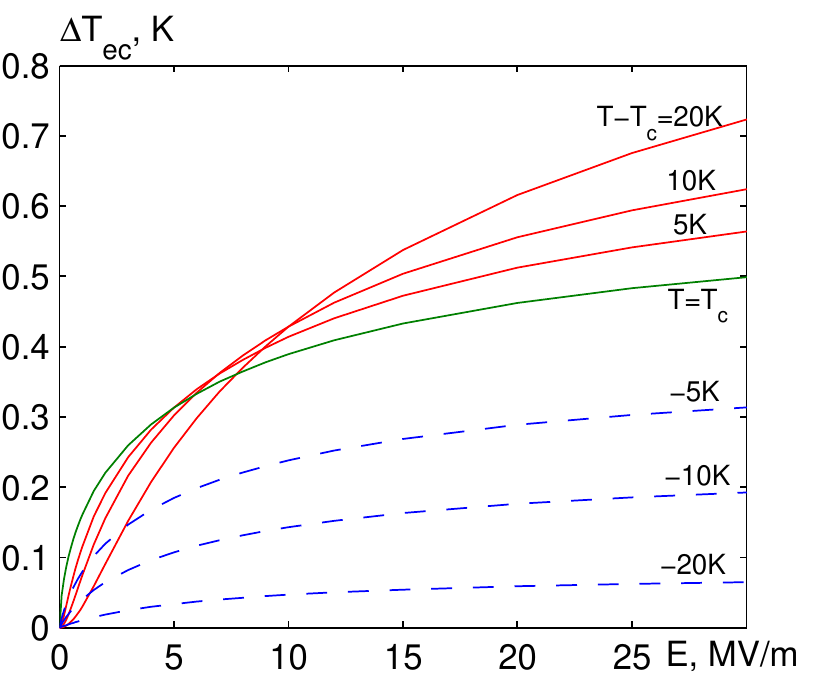} \includegraphics[scale=0.85]{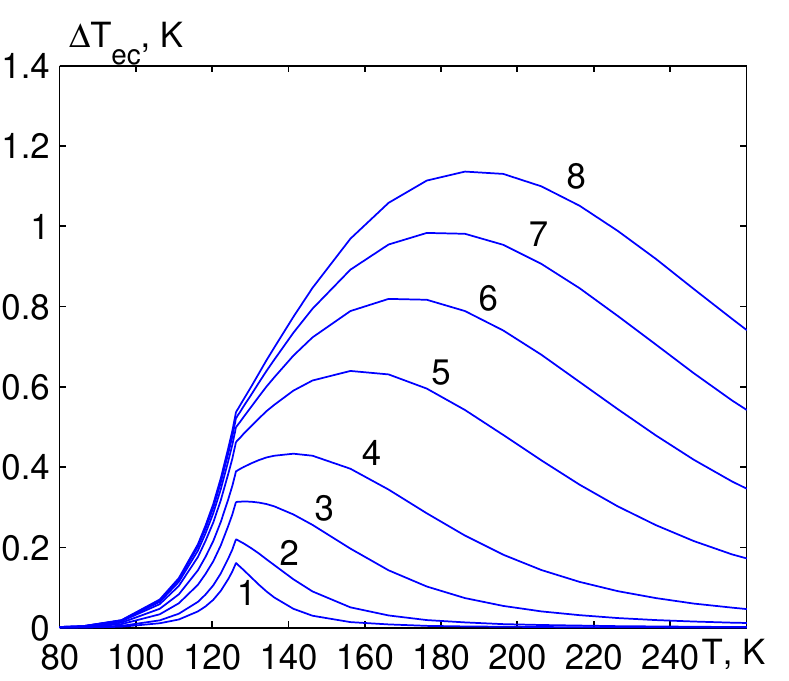} \\
		a ~~~~~~~~~~~~~~~~~~~~~~~~~~~~~~~~~~~~~~~~~~~~~~~~~~~~~~~~~~~~~~~ b
	\end{center}
	\caption{(Colour online) a) Field dependence of electrocaloric temperature change $\Delta T_{\text{ec}}$ at different temperature values $\Delta T = T-T_c$ and at hydrostatic pressure $p=0.3$~GPa. b) Temperature dependence of the electrocaloric temperature change $\Delta T_{\text{ec}}$ at different values of the longitudinal electric field $E_2$ (MV/m): 1.0 -- 1; 2.0 -- 2; 5.0 -- 3; 10.0 -- 4; 20.0 -- 5; 30.0 -- 6; 40.0 -- 7; 50.0 -- 8 and at hydrostatic pressure $p=0.3$~GPa.	 } 
	\label{DTec_E_3kbar}
\end{figure}

At high pressures, but less than the critical one, the field and temperature dependences of $\Delta T_{\text{ec}}$ are qualitatively similar, as in the absence of pressure (figure~\ref{DTec_E_3kbar}).

\begin{figure}[!h]
	\begin{center}
		\includegraphics[scale=0.85]{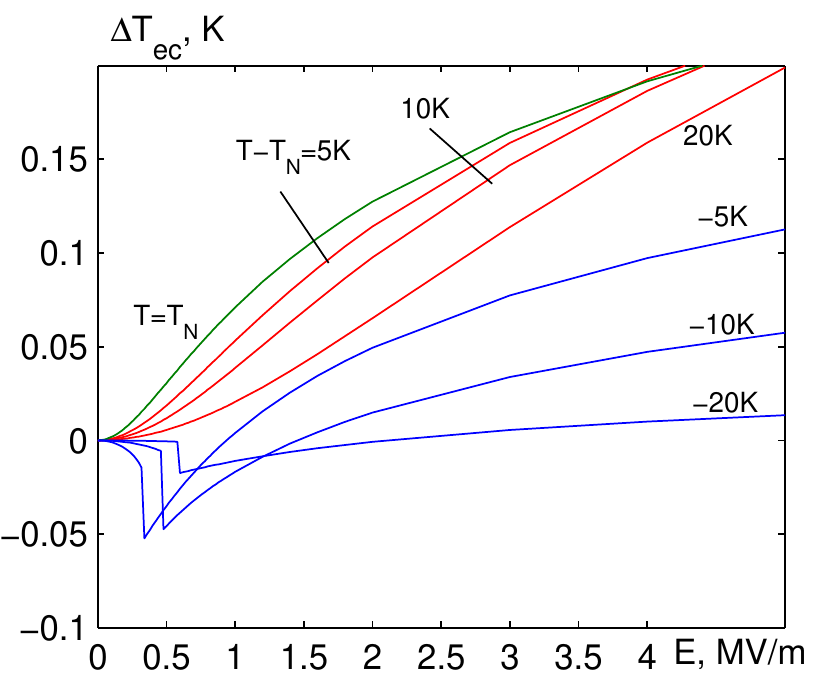} \includegraphics[scale=0.85]{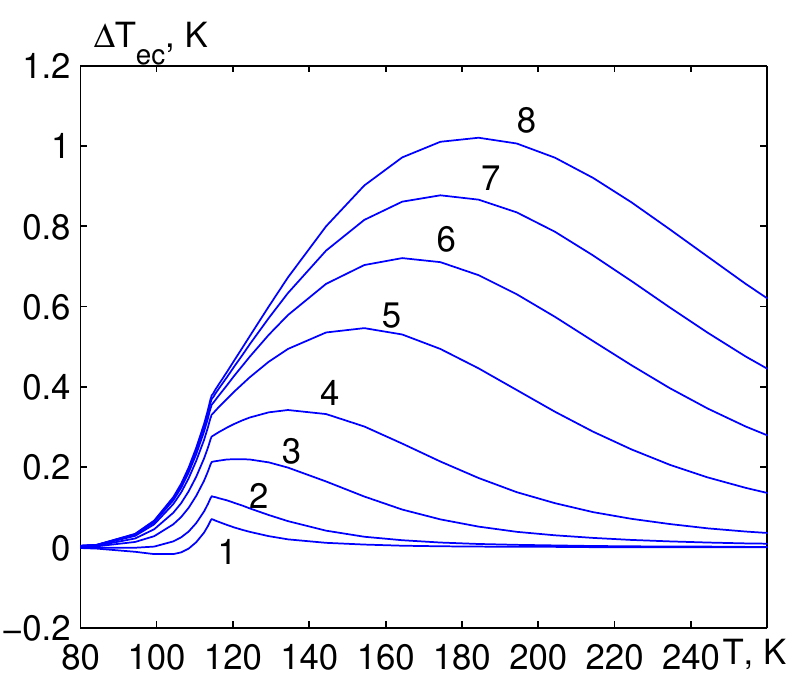} \\
		a ~~~~~~~~~~~~~~~~~~~~~~~~~~~~~~~~~~~~~~~~~~~~~~~~~~~~~~~~~~~~~~~ b
	\end{center}
	\caption[]{(Colour online) a) Field dependence of electrocaloric change of temperature $\Delta T_{\text{ec}}$ at different values of initial temperature $\Delta T = T-T_c$ and at hydrostatic pressure $p=0.45$~GPa. b) Temperature dependence of  $\Delta T_{\text{ec}}$ at different values of the longitudinal electric field $E_2$ (MV/m): 1.0 -- 1; 2.0 -- 2; 5.0 -- 3; 10.0 -- 4; 20.0 -- 5; 30.0 -- 6; 40.0 -- 7; 50.0 -- 8 and at hydrostatic pressure $p=0.5$~GPa. } \label{DTec_E_4_5kbar}
\end{figure}

At pressures greater than the critical one, at temperatures $T\geqslant T_N$, EC effect is qualitatively similar to the case of subcritical pressures in the paraelectric phase: at weak fields $\Delta T_{\text{ec}}\sim E_2^2$ (green and red curves in figure~ \ref{DTec_E_4_5kbar}a), at strong fields, the $\Delta T_{\text{ec}}(E_2)$ dependencies deviate from the quadratic law.
At initial temperatures $T<T_N$ and weak fields $E_2$, the temperature of the crystal decreases nonlinearly with the field (blue curves in figure~\ref{DTec_E_4_5kbar}a).
This is due to antiferroelectric ordering because the crystal passes into the antiferroelectric phase at pressures higher than the critical one. The ordering of pseudospins in sublattice ``B'' (which is oriented opposite to the field) under the action of the field is stronger than the ordering of pseudospins in sublattice ``A'', which leads to the isothermal increase of entropy and adiabatic (at constant entropy) lowering of temperature.
With the further strengthening of the field, the pseudospins in the ``B'' sublattice are overturned and ordered in the direction of the field, which leads to the isothermal decrease of entropy and to the isoentropic increase of temperature.

\begin{figure}[!t]
	\begin{center}
		\includegraphics[scale=0.75]{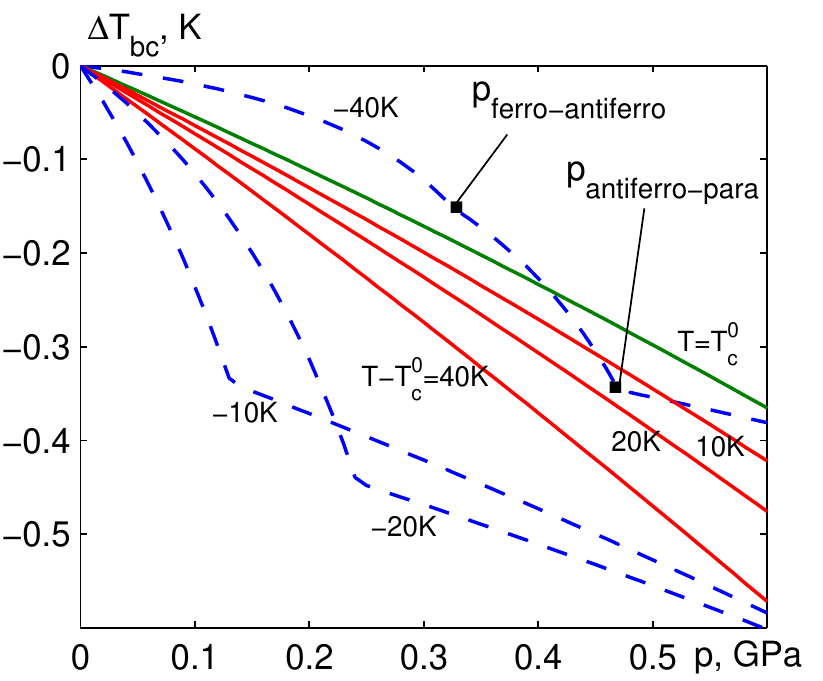} \includegraphics[scale=0.75]{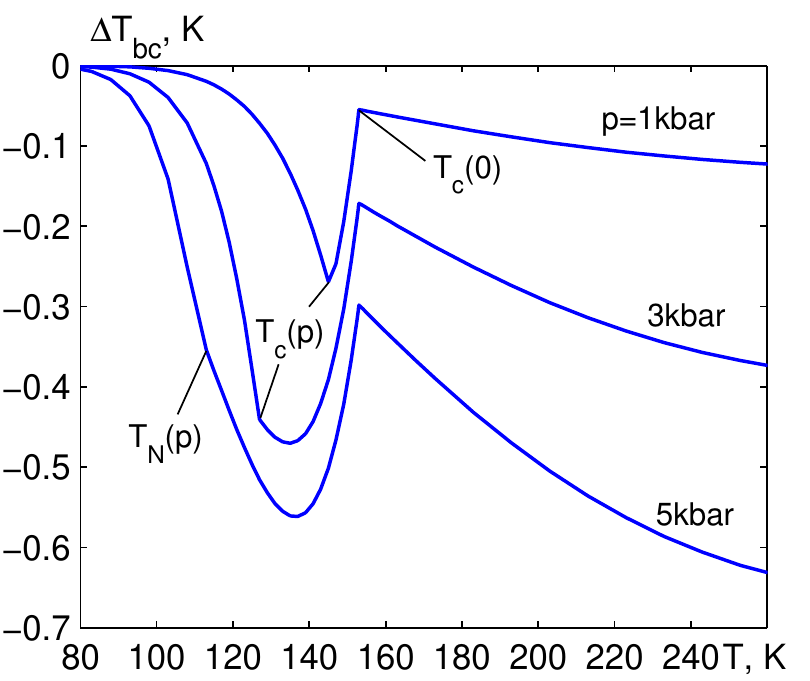} \\
		a ~~~~~~~~~~~~~~~~~~~~~~~~~~~~~~~~~~~~~~~~~~~~~~~~~~~~~~~~~~~~~~~ b
	\end{center}
	\caption{(Colour online) a) Pressure dependence of the barocaloric temperature change $\Delta T_{\text{bc}}$ at different values of temperature $\Delta T = T-T_c^0$ and in the absence of a field.   b) Temperature dependence of barocaloric temperature change $\Delta T_{\text{bc}}$ at different values of adiabatically applied pressure $p$ and in the absence of a field.
	} \label{DTbc_p}
\end{figure}

Hydrostatic pressure $p$ lowers the Curie temperature. This leads to the isothermal increase of entropy and to the isentropic lowering of temperature, as shown in figure~\ref{S_E_p}. Therefore $\Delta T_{\text{bc}}$ is negative and at $T \geqslant T_c^0$ it lowers almost linearly with increasing pressure (figure~\ref{DTbc_p}a, green and red solid curves).
At $T<T_c^0$ (ferroelectric phase) at low pressures, the BC effect is stronger than in the paraelectric phase (in figure~\ref{DTbc_p}a these are the blue dashed curves corresponding to $T-T_c^0=-10K,\,\, -20$~K). At a certain value of pressure, the crystal passes to the paraelectric phase (see figure~\ref{TcTN_E2}), in which the rate of cooling with pressure is less, and therefore a break appears in the $\Delta T_{\text{bc}}(p)$ curve.
\begin{figure}[!t]
	\begin{center}
		\includegraphics[scale=0.7]{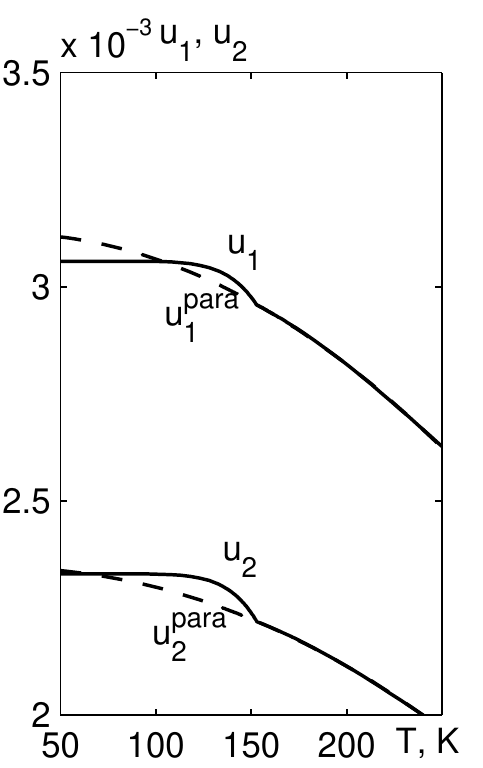} ~~~~ \includegraphics[scale=0.7]{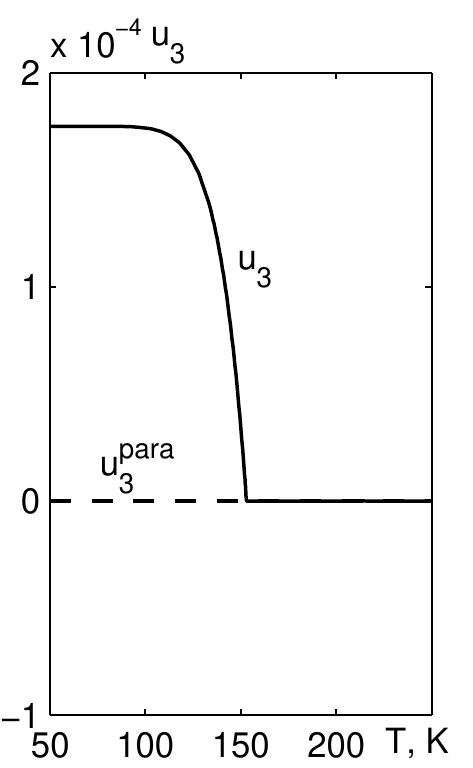}~~~~ \includegraphics[scale=0.7]{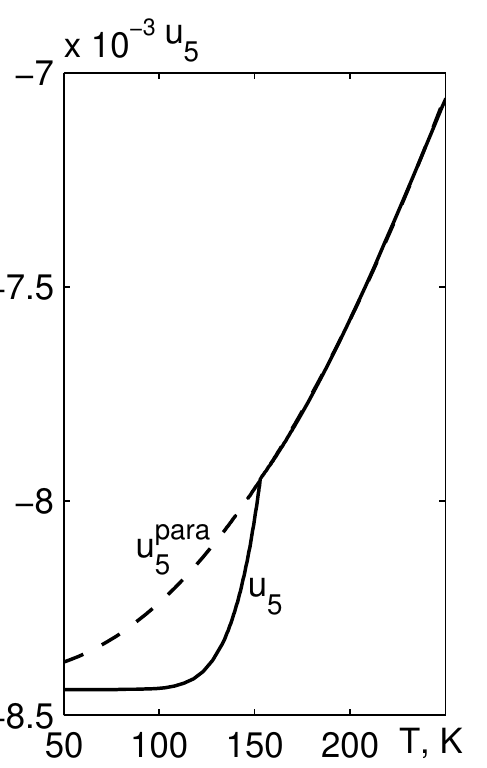}\\
	\end{center}
	\caption{Temperature dependence of lattice  strains  $u_j$ under zero pressure, calculated in \cite{Vdovych33702}.} \label{u_j}
\end{figure}

As can be seen from figure~\ref{TcTN_E2}, at $T-T_c^0=-40$~K there are two phase transitions when increasing pressure: from ferroelectric to antiferroelectric phase, and then from antiferroelectric to paraelectric phase. Accordingly, in figure~\ref{DTbc_p}a two breaks appear on the curve $\Delta T_{\text{bc}} (p)$.

It should be noted that in this work, only the pseudospin (proton) contribution to the BC effect was calculated, and lattice anharmonicities were not taken into account. The interaction between pseudospins leads to the occurrence of stretching strains due to the electrostrictive coupling of the pseudospin and lattice subsystems, since after substitution of (\ref{w}) into (\ref{Hshort}) and also (\ref{nu}) into (\ref {H0}),  there appear terms of the type $\delta_{j}u_j\frac{\sigma_{q1}^{A}}{2} \frac{\sigma_{q'2}^{A}}{2} $ and $\psi_{j1}u_j\eta_{1}^2$. The mean values of pseudospins decrease with an increase of temperature. As a result, the electrostrictive coupling becomes weaker and the diagonal strains $u_1$, $u_2$, $u_3$ decrease (figure~\ref{u_j}).

The volume of the crystal decreases along with strains, $(\partial V/\partial T)_{p}<0$. Therefore, $\Delta T_{\text{bc}}$ is negative, according to the formula (\ref{DTbc_int}).
We also note that it is possible to take a set of deformation potentials $\delta_{j}$ [see (\ref{w})] and $\psi_{j1}$, $\psi_{j2}$ [see (\ref{nu})], which leads to an increase of the volume of the crystal with an increase of temperature. However, this simultaneously leads to an increase in the Curie temperature with an increase in pressure, which contradicts the experimental data.

In contrast to electrostrictive coupling, lattice anharmonicities  lead to thermal expansion of the crystal and give a positive contribution to the BC effect. This contribution competes with the pseudospin contribution, and, in a certain temperature range, it can be larger than the pseudospin contribution.

\section{Conclusions}

In the case of a weak longitudinal field $E_2$, the electrocaloric change in temperature $\Delta T_{\text{ec}}$ increases linearly with the field in the ferroelectric phase, quadratically in the paraelectric phase, and according to the law $\Delta T_{\text{ec}}\sim E_2^{2/3}$ at the initial temperature $T=T_c$. In the strong field, the dependences $\Delta T_{\text{ec}}(E_2)$ deviate from the mentioned laws.
Applying the hydrostatic pressure, the EC effect is qualitatively similar to the one at zero pressure. At pressures greater than the critical one, the EC effect may be negative due to the transition of the crystal into the antiferroelectric phase.

The barocaloric change in temperature $\Delta T_{\text{bc}}$ has a negative sign and decreases almost linearly with pressure since the Curie temperature decreases with pressure. The nonlinearity is strongly manifested at low initial temperatures.
In our calculations, only the pseudospin contribution to the BC effect is taken into account. The electrostrictive coupling of the pseudospin and lattice subsystem leads to a decrease in the volume of the crystal with increasing temperature, and as a result  the BC effect is negative. To obtain~$\Delta T_{\text{bc}}$, which can be compared with experimental data, it is necessary to take into account the thermal expansion associated with the lattice anharmonicities.

\section*{Appendix. Notations in the expression for molar heat capacity}
\setcounter{equation}{0}
\renewcommand{\theequation}{A.\arabic{equation}}
The notations introduced in expression (\ref{DC}) are as follows:
\bea
%&& S'_{T} = \frac{R}{N_m} ( -2\eta_1 \frac{\beta}{T} (\nu_{1}\eta_{1}  +   \nu_{2} \eta_{2} + \frac{1}{2}(\mu_yE_2 + 3\eta_1^2 \mu'E_2)) -  \nonumber \\
%&&   ~~~~~~~~~~~~~~~  -2\eta_2 \frac{\beta}{T} (\nu_{2}\eta_{1}  +   \nu_{1} \eta_{2} + \frac{1}{2}(\mu_yE_2 + 3\eta_2^2 \mu'E_2)) + \frac{4M\beta w}{DT} ), \label{Sx}\\
&& \hspace{-8ex}  S'_{T} = \frac{R}{N_m} \left(  \frac{4\beta w}{D} \left[  y_1^T a\sinh y_1  + y_2^T a\sinh y_2   +  \frac{\beta w}{T}aM^a \right]   -  
\frac{4M\beta w}{D} \left[ y_1^T\eta_1 + y_2^T\eta_2 +\frac{\beta}{T} \frac{2M w}{D}\right]  \right) ,\nonumber \\
%
%&&\hspace{-8ex}  S'_{\eta_1} = \frac{R}{N_m} ( -\frac{2\eta_1}{ 1 - \eta_1^2} + 2[\eta_1y_1^{\eta_1} + \eta_2  \beta\nu_2] -  
%                                 4\beta \nu_{1}\eta_{1}  -  4\beta  \nu_{2} \eta_{2} - (\beta\mu_yE_2 + 9 \beta\eta_1^2 \mu'E_2)  +  \nonumber \\
%&&   ~~~~~~~~~~~~~~~~~~~ + \frac{4\beta w}{D} ( y_1^{\eta_1}a\sinh y_1 + \beta\nu_2a\sinh y_2) - \frac{4M\beta w}{D} [\eta_1y_1^{\eta_1} + \eta_2  \beta\nu_2]),  \nonumber \\
&&\hspace{-8ex}  S'_{\eta_1} = \frac{R}{N_m} \left(  2T y_1^T  +  \frac{4\beta w}{D} \left(  y_1^{\eta_1}a\sinh y_1 + \beta\nu_2a\sinh y_2\right)  - \frac{4M\beta w}{D} \left[ \eta_1y_1^{\eta_1} + \eta_2  \beta\nu_2\right] \right) ,  \nonumber \\
%
%&&\hspace{-8ex}  S'_{\eta_2} = \frac{R}{N_m} ( -\frac{2\eta_2}{ 1 - \eta_2^2} + 2 [\eta_2y_2^{\eta_2} + \eta_1  \beta\nu_2] - 4\beta \nu_{2}\eta_{1}  -  4\beta  \nu_{1} \eta_{2} - (\beta\mu_yE_2 + 9 \beta\eta_1^2 \mu'E_2)  + \nonumber \\
%&&   ~~~~~~~~~~~~~~~~~~~~~~~~~~~~~ + \frac{4\beta w}{D} ( \beta\nu_2a\sinh y_1 + y_2^{\eta_2}a\sinh y_2) - \frac{4M\beta w}{D} [\eta_2y_2^{\eta_2} + \eta_1  \beta\nu_2]),  \nonumber \\
&&\hspace{-8ex}  S'_{\eta_2} = \frac{R}{N_m} \left(  2T y_2^T  +  \frac{4\beta w}{D} \left(  \beta\nu_2a\sinh y_1 + y_2^{\eta_2}a\sinh y_2\right)  - \frac{4M\beta w}{D} \left[ \eta_2y_2^{\eta_2} + \eta_1  \beta\nu_2\right] \right) ,  \nonumber \\
%&& S'_{u_j} = \frac{R}{N_m} ( \frac{2}{D} [ D\eta_1(\beta \psi_{j1}\eta_1+\beta \psi_{j2}\eta_2) + D\eta_2(\beta \psi_{j2}\eta_1+\beta \psi_{j1}\eta_2) - 2M\beta\delta_j ] ), 
&&\hspace{-8ex}  S'_{u_j} = \frac{R}{N_m} \left(  \frac{4\beta w}{D} \left[  y_1^{u_j} a\sinh y_1   + y_2^{u_j} a\sinh y_2  -  \beta\delta_j a M^a \right]  -  \frac{4M\beta w}{D} \left[  \eta_1y_1^{u_j} + \eta_2y_2^{u_j} - \frac{2M\beta\delta_j}{D} \right]  \right) .\label{Sx} 
\eea
Here are the notations:          
\bea
&&  y_1^T = -\frac{\beta}{T} \left[ \nu_1\eta_1  + \nu_2\eta_2 + \frac{1}{2}\left( \mu_yE_2 + 3\eta_1^2 \mu'E_2\right) \right], ~~ 
y_2^T = -\frac{\beta}{T} \left[ \nu_2\eta_1  + \nu_1\eta_2 + \frac{1}{2}\left( \mu_yE_2 + 3\eta_2^2 \mu'E_2\right) \right] . \nonumber \\
&& y_1^{\eta_1} = \frac{1}{1-\eta_1^2} +  \beta\nu_1 + 3\beta\eta_1 \mu'E_2, ~~~~ y_2^{\eta_2} = \frac{1}{1-\eta_2^2} +  \beta\nu_1 + 3\beta\eta_2 \mu'E_2,  \nonumber \\
&& y_1^{u_j} = \beta(\psi_{j1}\eta_1+\psi_{j2}\eta_2), ~~~~ y_2^{u_j} = \beta(\psi_{j2}\eta_1+\psi_{j1}\eta_2),  \nonumber \\
&&\hspace{-2ex}\phantom{\frac{1}{2}} M^a = \cosh y_1 + \cosh y_2 + 4a. \nonumber 
\eea

After differentiating the system of equations (\ref{eta1eta2}) with
respect to the temperature, we obtain a system of equations, from which we determine $\eta'_{Tf}$ and $u'_{Tj}$:
\bea
&& \hspace{-10ex} \left( \!\begin{tabular}{cc}
	$\hat{A}^{\eta}-\hat{I}$  &  $\hat{A}^{u}$  \\ 
	$\hat{B}^{\eta}$  &  $\hat{B}^{u}$  \\ 
\end{tabular}  \!\right) 
\left( \!\begin{tabular}{c}
	$\vec{\eta}'_{T}$   \\ 
	$\vec{u}'_{T}$   \\ 
\end{tabular}  \!\right) + 
\left(\! \begin{tabular}{c}
	$\vec{\!A}^{T}$   \\ 
	$\vec{\!B}^{T}$   \\ 
\end{tabular}  \!\right) = \vec{0}.   ~ \Rightarrow ~  
\left(\! \begin{tabular}{c}
	$\vec{\eta}'_{T}$   \\ 
	$\vec{u}'_{T}$   \\ 
\end{tabular}  \!\right) = 
-\left(\! \begin{tabular}{cc}
	$\hat{A}^{\eta}-\hat{I}$  &  $\hat{A}^{u}$  \\ 
	$\hat{B}^{\eta}$  &  $\hat{B}^{u}$  \\ 
\end{tabular}  \!\right)^{-1} 
\left(\! \begin{tabular}{c}
	$\vec{\!A}^{T}$   \\ 
	$\vec{\!B}^{T}$   \\ 
\end{tabular}  \!\right),
\label{ABetaeps} 
\eea
where $\hat{I}$ is a 2$\times$2 identity matrix. Coefficients of the $\hat{A}^{\eta}$ matrix are:
\bea
&& \!\!\!\!A^{\eta}_{11}= \eta_1^{y_1} y_1^{\eta_1}  + \eta_1^{y_2} \beta\nu_2, ~~~ 
A^{\eta}_{12}= \eta_1^{y_1} \beta\nu_2  + \eta_1^{y_2} y_2^{\eta_2}, ~~ \nonumber \\
&& \!\!\!\!A^{\eta}_{21}= \eta_2^{y_1} y_1^{\eta_1}  + \eta_2^{y_2} \beta\nu_2, ~~~  
A^{\eta}_{22}= \eta_2^{y_1} \beta\nu_2  + \eta_2^{y_2} y_2^{\eta_2}, ~~ \nonumber 
\eea
where the notations are entered:             
\bea
&& \eta_1^{y_1}= \frac{1}{D} \left[  \cosh (y_1 + y_2) +  \cosh (y_1 - y_2) + 2a \cosh y_1  - \eta_1^2 \right] ,  \nonumber \\
&& \eta_1^{y_2}=  \eta_2^{y_1}= \frac{1}{D} \left[  \cosh (y_1 + y_2) -  \cosh (y_1 - y_2)   - \eta_1\eta_2 \right] , ~~ \nonumber \\
%&& \eta_2^{y_1}= \frac{1}{D} [ \cosh (y_1 + y_2) -  \cosh (y_1 - y_2)  - \eta_1\eta_2 ],  \nonumber \\
&& \eta_2^{y_2}= \frac{1}{D} \left[  \cosh (y_1 + y_2) +  \cosh (y_1 - y_2) + 2a \cosh y_2   - \eta_2^2 \right] , ~~ \nonumber 
\eea
coefficients of matrix $\hat{A}^{u}$:         
\bea
&& \!\!\!\!A^{u}_{1j}= \eta_1^{y_1} y_1^{u_j}  + \eta_1^{y_2} y_2^{u_j} - 
\frac{\beta\delta_j}{D} \left[   2a \sinh y_1 - 2M\eta_1\right] ,  \nonumber \\
&& \!\!\!\!A^{u}_{2j}= \eta_2^{y_1} y_1^{u_j}  + \eta_2^{y_2} y_2^{u_j} - 
\frac{\beta\delta_j}{D} \left[   2a \sinh y_2 - 2M\eta_2\right]  , ~ \nonumber 
\eea
coefficients of matrix $\hat{B}^{\eta}$:         
\bea
&& \hspace{-4ex} B^{\eta}_{j1} = -\frac{2}{v} (\psi_{j1}\eta_1 + \psi_{j2}\eta_2) + \frac{4\delta_j}{vD}( a\sinh y_{1}y_1^{\eta_1} 
+ a\sinh y_{2}  \beta\nu_2)  - \frac{4M\delta_j}{vD}\left(  \eta_1y_1^{\eta_1} + \eta_2  \beta\nu_2\right) , \label{Beta1} \nonumber\\
&& \hspace{-4ex} B^{\eta}_{j2} = -\frac{2}{v} (\psi_{j1}\eta_2 + \psi_{j2}\eta_1) + \frac{4\delta_j}{vD}\left(  a\sinh y_{1} \beta\nu_2 
+ a\sinh y_{2} y_2^{\eta_2} \right)   - \frac{4M\delta_j}{vD}( \eta_1 \beta\nu_2 + \eta_2 y_2^{\eta_2}), \label{Beta2} \nonumber
\eea
coefficients of matrix  $\hat{B}^{u}$:         
\bea
&& \hspace{-4ex} B^{u}_{jj'} = c_{jj'}^{E0}  + \frac{4\delta_j}{vD} \left[  y_1^{u_{j'}} a\sinh y_1   + y_2^{u_{j'}} a\sinh y_2  -  \beta\delta_{j'} a M^a \right]   - \frac{4M\delta_j}{vD} \left[  \eta_1 y_1^{u_{j'}} + \eta_2 y_2^{u_{j'}} - \frac{2M\beta\delta_{j'}}{D} \right] , \label{Bu} \nonumber
\eea
coefficients of vectors $\vec{\!A}^{T}$  and  $\vec{\!B}^{T}$:        
\bea
&& A^{T}_1 = \eta_1^{y_1} y_1^T  + \eta_1^{y_2} y_2^T + \frac{\beta w}{DT} \left( 2a \sinh y_1 - 2M\eta_1\right),  \nonumber \\
&& A^{T}_2 = \eta_2^{y_1} y_1^T  + \eta_2^{y_2} y_2^T + \frac{\beta w}{DT} \left( 2a \sinh y_2 - 2M\eta_2\right) ,  \nonumber \\
&& B^{T}_j = \frac{4\delta_j}{vD} \left[  y_1^T a\sinh y_1  + y_2^T a\sinh y_2   +  \frac{a M^a \beta w}{T} \right]   -  \frac{4\delta_jM}{vD} \left[ y_1^T\eta_1 + y_2^T\eta_2 +  \frac{2M \beta w}{DT}\right] . \nonumber 
\eea

\ukrainianpart

\title{Електрокалоричний і барокалоричний  ефекти у сегнетоелектрику  CsH$_2$PO$_4$}
\author{А. С. Вдович \refaddr{label1}, \framebox{ Р. Р. Левицький}\refaddr{label1}, І. Р. Зачек \refaddr{label2}}
\addresses{
	\addr{label1} Інститут фізики конденсованих систем Національної академії наук України, \\вул. Свєнціцького, 1, 79011 Львів, Україна
	\addr{label2} Національний університет ``Львівська політехніка'', Україна, 79013, Львів, вул.~С.~Бандери,  12}

\makeukrtitle

\begin{abstract}
	\tolerance=3000%
	Для дослідження калоричних ефектів у сегнетоелектрику  CsH$_2$PO$_4$  використано модифіковану псевдоспінову модель цього кристала, яка
	враховує залежність параметрів взаємодії між псевдоспінами від деформацій гратки. Модель також враховує залежність ефективного дипольного момента на водневому зв'язку  від параметра впорядкування. В наближенні двочастинкового кластера  вивчено вплив  поздовжнього електричного поля і гідростатичного тиску на  молярну ентропію кристала.  Досліджено  електро\-кало\-рич\-ний і барокалоричний  ефекти.  Розрахована електрокалорична зміна температури  близько 1~K; вона може міняти знак під дією гідростатичного тиску. Барокалорична зміна температури близько $-0.5$~K; при її розрахунках не враховувалися ангармонізми гратки.
	\keywords сегнетоелектрики, сегнетоелектричний фазовий перехід, електрокалоричний ефект, барокалоричний ефект
	
\end{abstract}


\begin{thebibliography}{99}


\bibitem{Mischenko2006} Mischenko A. S., Zhang Q., Scott J. F., Whatmore R. W., Mathur N. D., Science,  2006, \textbf{311}, No.~5765, 1270--1271, \doi{10.1126/science.1123811}.

\bibitem{Peng2987} Peng B., Fan H.,  Zhang Q.,   Adv. Funct. Mater., 2013,  \textbf{23}, No.~23, 2987--2992, \doi{10.1002/adfm.201202525}.

\bibitem{Peng1708} Peng B., Zhang Q.,  Gang B., Leighton G. J. T., Shaw C., Milne S. J., Zou B., Sun W., Huang H., Wang Z.,   Energy Environ. Sci., 2019, \textbf{12}, No.~5, 1708--1717,  \doi{10.1039/c9ee00269c}.

\bibitem{Lu162904}  Lu S. G., Rozic B., Zhang Q. M.,  Kutnjak Z.,  Li X.,  Furman E.,  Gorny L. J., Lin M.,  Malic B.,  Kosec M., \\ Blinc R.,   Pirc R.,   
Appl. Phys. Lett., 2010, \textbf{97}, No.~16, 162904 (3 pages), \doi{10.1063/1.3501975}.


\bibitem{Asbani164517} Asbani B., Dellis J. L., Lahmar A., Amjoud M., Gagou Y., Mezzane D., Kutnjak Z., Pirc R., El Marssi M., Luk’yanchuk I., Rozic B.,  J. Alloys Compd., 2022, \textbf{907}, 164517 (5 pages),  \doi{10.1016/j.jallcom.2022.164517}. 

\bibitem{Nouchokgwe114873} Nouchokgwe Y., Lheritier P., Usui T., Torello A., El Moul A., Kovacova V., Granzow T., Hirose S., Defay E.,   Scr. Mater., 2022, \textbf{219}, 114873 (5 pages),  \doi{10.1016/j.scriptamat.2022.114873}.

\bibitem{Liu2021} Liu X., Wu Z., Guan T., Jiang H., Long P., Li X., Ji C., Chen S., Sun Z., Luo J.,   Nat. Commun., 2021, \textbf{12}, \\5502 (7 pages), \doi{10.1038/s41467-021-25644-x}.

\bibitem{363x} Wiseman G. G., IEEE Trans. Electron Devices, 1969, \textbf{ED-16}, No.~6, 588--593, \doi{10.1109/T-ED.1969.16804}.

\bibitem{Baumgartner1950} Baumgartner H., Helv. Phys. Acta, 1950, \textbf{23}, 651--696, \doi{10.5169/seals-112128}.

\bibitem{Shimshoni1970}  Shimshoni M., Harnik E., J. Phys. Chem. Solids, 1970, \textbf{31}, No.~6, 1416--1417, \\\doi{10.1016/0022-3697(70)90148-4}.


\bibitem{Vdovych_CMP2014_el} Vdovych A. S., Moina A. P., Levitskii R. R., Zachek I. R.,  Condens. Matter Phys., 2014, \textbf{17}, No.~4, 43703 \\(10 pages),  \doi{10.5488/CMP.17.43703}. 

\bibitem{Meng2021} Meng Y., Pu J.,  Pei Q.,  Joule, 2021, \textbf{5}, No.~4, 780--793,  \doi{10.1016/j.joule.2020.12.018}.


\bibitem{Gorev2019} Gorev M. V., Mikhaleva E. A., Flerov I. N., Bogdanov E. V., J. Alloys Compd., 2019, \textbf{806}, 1047--1051,  \doi{10.1016/j.jallcom.2019.07.273}.

\bibitem{Lloveras2015} %Lloveras P., Stern-Taulats E., Barrio M. et al. 
Lloveras P., Stern-Taulats E., Barrio M., Tamarit J. Ll., Crossley S., Li W., Pomjakushin V., Planes A., \\Manosa Ll., Mathur N.D., Moya X., Nat. Commun., 2015, \textbf{6}, 8801 (6 pages), \doi{10.1038/ncomms9801}.
	
\bibitem{Vdovych33702}  Vdovych A. S.,  Zachek I. R.,  Levitskii R. R.,  Condens. Matter Phys., 2020,  \textbf{23}, No.~3, 33702 (16 pages), \doi{10.5488/CMP.23.33702}.
	
\bibitem{Matsunaga2011} Matsunaga H., Itoh K., Nakamura E., J. Phys. Soc. Jpn., 1980, \textbf{48}, No.~6, 2011--2014, \doi{10.1143/JPSJ.48.2011}.  

\bibitem{Itoh2626} Itoh K., Hagiwara T., Nakamura E., J. Phys. Soc. Jpn., 1983, \textbf{52}, No.~8, 2626--2629, \doi{10.1143/JPSJ.52.2626}.       

\bibitem{Iwata304} Iwata Y., Koyano N., Shibuya I., J. Phys. Soc. Jpn., 1980, \textbf{49}, No.~1, 304--307, \doi{10.1143/JPSJ.49.304}.

\bibitem{Iwata4044} Iwata Y., Deguchi K., Mitani S., Shibuya I., Onodera Y., Nakamura E., J. Phys. Soc. Jpn., 1994, \textbf{63}, No.~11, 4044--4050, \doi{10.1143/JPSJ.63.4044}.

\bibitem{Yasuda1311} Yasuda N., Okamoto M., Shimizu H., Fujimoto S., Yoshino K., Inuishi Y., Phys. Rev. Lett., 1978, \textbf{41}, No.~19, 1311--1314, \doi{10.1103/PhysRevLett.41.1311}.

\bibitem{Yasuda2755} Yasuda N., Fujimoto S., Okamoto M., Shimizu H., Yoshino K., Inuishi Y.,  Phys. Rev. B, 1979, \textbf{20}, No.~7, 2755--2764, \doi{10.1103/PhysRevB.20.2755}.

\bibitem{Schuele935} Schuele P. J.,  Thoma R. A., Jpn. J. Appl. Phys., 1985, \textbf{24}, No.~S2, 935--937,  \doi{10.7567/JJAPS.24S2.935}.

\bibitem{Schuele2549} Schuele P. J., Schmidt V. H.,   Phys. Rev. B, 1989, \textbf{39}, No.~4, 2549--2556,  \doi{10.1103/PhysRevB.39.2549}.

\bibitem{Kobayashi83} Kobayashi Yu., Deguchi K., Azuma Sh., Suzuki E., Ming Li Ch., Endo Sh., Kikegawad T., Ferroelectrics, 2003, \textbf{285}, No.~7, 83--89,  \doi{10.1080/00150190390205924}.

\bibitem{Deguchi024106}  Deguchi K., Azuma S., Kobayashi Y., Endo S., Tokunaga M.,    Phys. Rev. B, 2004, \textbf{69}, No.~2, 024106, \\ \doi{10.1103/PhysRevB.69.024106}.
	
\bibitem{Gesi} Gesi K., Ozawa K.,  Jpn. J. Appl. Phys., 1978,  \textbf{17}, No.~2, 435--436, \doi{10.1143/JJAP.17.435}.

\bibitem{Brandt} Brandt N.  B., Zhukov S. G., Kulbachinskii V. A., Smirnov P. S., Strukov B. A., Fiz. Tverd. Tela, 1986, \textbf{28}, 3159, (in Russian).


\bibitem{Magome2010} Magome E., Tomiaka S., Tao Y., Komukae M., J. Phys. Soc Jpn., 2010, \textbf{79}, No.~2,  025002 (2 pages),  \doi{10.1143/JPSJ.79.025002}.
	
\bibitem{Imai3960} Imai K., J. Phys. Soc. Jpn., 1983, \textbf{52}, No.~11, 3960--3965, \doi{10.1143/JPSJ.52.3960}.

\bibitem{Shchur054301} Shchur Ya.,   Phys. Rev. B, 2006,  \textbf{74},  054301 (8 pages), \doi{10.1103/PhysRevB.74.054301}.

\bibitem{Shchur569} Shchur Ya.,  Phys. Status Solidi B,  2007,  \textbf{244}, No.~2,  569--577, 	\doi{10.1002/pssb.200642176}.
	
\bibitem{Lasave134112} Lasave J., Abufager P., Koval S., Phys. Rev. B, 2016, \textbf{93}, No.~13, 134112 (11 pages), \\\doi{10.1103/PhysRevB.93.134112}.

\bibitem{Kojyo4391} Kojyo N.,  Onodera Y.,  J. Phys. Soc. Jpn. 1988, \textbf{57}, No.~12,  4391--4402,  	\doi{10.1143/JPSJ.57.4391}.


\bibitem{Shchur301} Shchur Ya., Bryk T., Klevets I., Kityk A. V.,  Comput. Mater. Sci., 2016, \textbf{111}, 301--309, \\\doi{10.1016/j.commatsci.2015.09.014}.

\bibitem{VanTroeye024112} Van Troeye B., van Setten M. J., Giantomassi M., Torrent M., Rignanese G.-M.,  Gonze X.,  \\Phys. Rev. B, 2017, \textbf{95}, No.~2, 024112 (9 pages),  \doi{10.1103/PhysRevB.95.024112}.
	
\bibitem{Blinc6031} Blinc R., SaBaretto F. C., J. Chem. Phys., 1980, \textbf{72}, No.~11, 6031--6034, \doi{10.1063/1.439058}.

\bibitem{914R} Stasyuk I. V., Levytsky R. R., Zachek I. R.,  Shchur Ya. Y., Kutny J. V., Miz E. V.,  Preprint \\of the Institute for Condensed Matter Physics, ICMP–91–4R, Lviv,
1991, (in Russian).
	
\bibitem{Deguchi3074} Deguchi K., Okaue E., Ushio S., Nakamura E., Abe K., J. Phys. Soc. Jpn., 1984, \textbf{53}, No.~9, \\3074--3080, \doi{10.1143/JPSJ.53.3074}.

\bibitem{FXTT40} Levitskii R. R., Zachek I. R., Vdovych A. S., J. Phys. Chem. Solids, 2012, \textbf{13}, No.~1, 40--47.
\newpage

\bibitem{Levitskii4702} Levitskii R. R., Zachek I. R., Vdovych A. S., J. Phys. Stud., 2012, \textbf{16}, No.~4, 4702 (11 pages), \\\doi{10.30970/jps.16.4702} (in Ukrainian).

\bibitem{Vdovych_JPS2021_CDP}   Vdovych	A. S., Levitskii R. R., Zachek I. R., Moina A. P.,  J. Phys. Stud., 2021,  \textbf{25}, No.~3, 3702 (14 pages),  
\doi{10.30970/jps.25.3702}.
	
\bibitem{Prawer63} Prawer S., Smith T. F.,  Finlayson T. R.,  Aust. J. Phys., 1985, \textbf{38}, No.~1, 63--83,  \doi{10.1071/PH850063}.


	
	
\end{thebibliography}
\end{document}